\documentclass[preprint2]{aastex}
\usepackage[]{natbib}

\begin{document}

\title{Neutrino emission from high-energy component gamma-ray bursts}

\author{Julia K.\ Becker} 
\affil{Fakult\"at f.\ Phys.\ \& Astron., Ruhr-Univ.\ Bochum, D-44780 Bochum, Germany}
\author{Francis Halzen}
\affil{Department of Physics, University of Wisconsin, Madison, WI-53706, USA}
\author{Aongus \'O Murchadha}
\affil{Department of Physics, University of Wisconsin, Madison, WI-53706, USA}
\author{Martino Olivo}
\affil{Fakult\"at f.\ Phys.\ \& Astron., Ruhr-Univ.\ Bochum, D-44780 Bochum, Germany}

\begin{abstract}
Gamma-ray bursts have the potential to produce the particle energies (up to $10^{21}$\,eV) and the energy budget ($10^{44}\, \rm{erg\, yr^{-1}\, Mpc^{-3}}$) to accommodate the spectrum of the highest energy cosmic rays; on the other hand, there is no observational evidence yet that they accelerate hadrons. Fermi recently observed two bursts that exhibit a power-law high-energy extension of the typical (Band) spectrum that extends to $\sim 30$ GeV. On the basis of fireball phenomenology we argue that they, along with GRB941017 observed by EGRET in 1994, show indirect evidence for considerable baryon loading. Since the detection of neutrinos is the only unambiguous way to establish that GRBs accelerate cosmic rays, we use two methods to estimate the neutrino flux produced when the baryons interact with fireball photons to produce charged pions and neutrinos. While the number of events expected from the Fermi bursts is small, we conclude that an event like GRB941017 will be detected by IceCube if gamma-ray bursts are indeed the sources of the cosmic rays. 
\end{abstract}


\section{Introduction \label{introduction}}

The sources of the extragalactic cosmic rays with energies in excess of $\sim$\,3$\times$10$^{18}$\,eV remain a mystery, but one of the best motivated candidates is gamma-ray bursts (GRBs). Large cosmic-ray energies can be achieved in the prompt phase of the GRB fireball where internal shocks have the potential to accelerate charged particles up to $\sim$10\,$^{21}$\,eV (\citealt{vietri_1995}, \citealt{waxman_1995}). Additionally, the total energy density in the Universe of cosmic rays must be matched by a sufficiently high hadronic energy density in the GRB with $\rho_{\mbox{\tiny{had,\,GRB}}}\approx\rho_{\mbox{\tiny{CR}}}$. GRB observations identify synchrotron photons produced by the electrons accelerated in the fireball with an energy $\epsilon_e\mbox{E}_{\mbox{\tiny{TOT}}}\sim\,$10$^{53}$\,ergs, where E$_{\mbox{\tiny{TOT}}}$ is the total energy released by the burst. GRB fireballs also carry energy $\epsilon_B\mbox{E}_{\mbox{\tiny{TOT}}}$ in the form of magnetic fields, and, if they are the sources of cosmic rays, energy $\epsilon_p\mbox{E}_{\mbox{\tiny{TOT}}}$ in protons. Assuming equipartition, $\epsilon_B\simeq\,\epsilon_e$ and $\epsilon_e+\epsilon_p+\epsilon_B=1$.

GRBs emerge as credible sources for the ultra high-energy cosmic rays because their observed flux can be accommodated with an energy density in protons that is similar to that in electrons, or $\epsilon_p\,\simeq\,\epsilon_e$. Recent estimates of the local rate of GRBs yield a maximum of $\dot{n}_{0}\!\!\sim\!\! 1$~Gpc$^{-3}$~yr$^{-1}$ assuming that GRBs follow the star formation rate. For a stronger evolution with redshift, the local rate can be as low as $0.05$~Gpc$^{-3}$~yr$^{-1}$. Using this result we estimate the electromagnetic energy density from GRBs to be in the range
$\rho_{\mbox{\tiny{em,\,GRB}}}\approx\dot{\mbox{n}}_{0}\,\epsilon_e\,\mbox{E}_{\mbox{\tiny{TOT}}}=\mbox{5}\times\mbox{10}^{42}\mbox{\,-\,}10^{44}\,\,\mbox{ergs\,Mpc}^{-3}\,\mbox{yr}^{-1}$. In order for GRBs to be the sources of cosmic rays, their hadronic energy density $\rho_{had,GRB}$ needs to produce the observed cosmic-ray energy density $\rho_{\mbox{\tiny{had,\,GRB}}}=\dot{\mbox{n}}_{0}\,\epsilon_p\,E_{\mbox{\tiny{TOT}}}\approx\,10^{44}\,\mbox{ergs\,Mpc}^{-3}\,\mbox{yr}^{-1}$. We therefore conclude that $\epsilon_p/\epsilon_e\approx\,1\mbox{\,-\,}20$. The extragalactic spectrum is actually expected to extend to energies below the knee, but remains unobserved due to the larger contribution of galactic cosmic rays, see e.g. \citet{ahlers}. If this is the case, the average fraction of proton to electron energy needs to be larger.

Fermi recently observed two bursts, GRB090510 and GRB090902b, that show a statistically significant deviation from the typical GRB spectrum described by the Band function\,\citep{090510,090902b}. A flux of high energy events is detected that extends to energies $\sim$30\,GeV following a power-law spectrum (Table~\ref{tab:latparms}). This is similar to the potentially much more luminous burst observed by EGRET in 1994, GRB941017\,\citep{egret94}. In this paper we will investigate the possibility that the high energy component results from $\pi^{0}$-decays and is therefore a signature of proton acceleration.

\begin{table}[h!b!p!]
\caption{Spectral parameters of Fermi GRBs with power-law components. The total fluence in gamma rays $F_{\gamma}^{\mbox{\tiny{TOT}}}$ is the sum of the Band fluence $F_{\gamma}^{\mbox{\tiny{B}}}$ and the power-law fluence $F_{\gamma}^{\mbox{\tiny{HE}}}$. The fluence for GRB900510 is calculated over the energy range 10 keV--30 GeV, and the fluence for GRB09092b is calculated over the range 10 keV--10 GeV. The parameters for GRB941017 are tabulated in~\citet{egret94}.}
\begin{center}
\begin{tabular*}{1.0\columnwidth}{c|cc}
           & \small{GRB090510} & \small{GRB090902b} \\[3pt] \hline \hline
\small{z}  & \small{0.903} & \small{1.822} \\
$T_{90}$   & \small{2.1\,s} & \small{21.9\,s}  \\
$F_{\gamma}^{\mbox{\tiny{TOT}}}$ & \small{$5.02\times10^{-5}$\,ergs\,cm$^{-2}$} & \small{$4.36\times10^{-4}$\,ergs\,cm$^{-2}$}  \\
$F_{\gamma}^{\mbox{\tiny{HE}}}$ & \small{$1.84\times10^{-5}$\,ergs\,cm$^{-2}$} & \small{$1.05\times10^{-4}$\,ergs\,cm$^{-2}$}  \\
$\alpha_{\gamma}$     & \small{0.58} & \small{0.61}  \\
$\beta_{\gamma}$      & \small{2.83} & \small{3.80}  \\
$E_{0}$  & \small{2771\,keV} & \small{726\,keV}  \\
$\Gamma_{\gamma}$     & \small{1.62} & \small{1.93}  \\
$E_{\mbox{\tiny{MAX}}}$ & \small{30.53\,GeV} & \small{33.40\,GeV} \\
$\Gamma$     & \small{1260} & \small{1000}  \\
\end{tabular*}
\end{center}
\label{tab:latparms}
\end{table}

The main contribution to the initial opacity of the fireball comes from the annihilation of photons into $e^{\pm}$ pairs. The Fermi observation of a non-thermal spectrum up to an energy $E_{\mbox{\tiny{MAX}}}$ of tens of GeV can be used to constrain the minimum bulk Lorentz factor $\Gamma_{\mbox{\tiny{MIN}}}$ required to make the source optically thin at the time of the gamma-ray display. For all photons with energy $E\leq E_{\mbox{\tiny{MAX}}}$ the condition $\tau_{\gamma\gamma}(E)<1$
must be fulfilled where $\tau_{\gamma\gamma}$ is the opacity. The observation of photons with energies of tens of GeV requires highly relativistic outflows with $\Gamma \simeq 10^3$. Because, on the other hand, the observed energy flux of order $10^{-4}$\,erg/s is typical of an average burst, the large boost factor implies that the photon density in the rest frame of the burst is low. This is a strong effect as the photon density is suppressed by $\Gamma^{-4}$. From $\tau_{\gamma\gamma}(E)=1$, we can determine $\epsilon_{e}$ by finding the electromagnetic energy over the volume of the fireball as a fraction of the total GRB energy:
\begin{eqnarray}
\label{eq:epse}
\epsilon_{e}\!\!&\approx&\!\!1-5 \times 10^{-2} \left( \frac{\Gamma}{300}\right)^{4} \left( \frac{\delta t}{10 \,\rm{ms}}\right) \left( \frac{E_{\gamma}}{1 \,\rm{MeV}}\right)\nonumber\\
& & \times \left( \frac{T_{90}}{100\, \rm{s}}\right) \left( \frac{10^{53}\, \rm{ergs}}{E_{\mbox{\tiny TOT}}}\right)
\end{eqnarray}
where $\Gamma$ is the bulk Lorentz factor of the fireball, $\delta t$ is the variability timescale, $E_{\gamma}$ is the characteristic gamma-ray energy (which we take to be the peak energy of the event), $T_{90}$ is the duration of the burst, and $E_{\mbox{\tiny{TOT}}}$ is the total energy of the burst. We therefore estimate $\epsilon_{e}$ for GRB090510 to be $\sim0.05$ and for GRB090902b to be $\sim0.02$. From the low values of $\epsilon_e$ thus obtained we must conclude that protons dominate the fireball in order to accommodate the total energy of a solar mass generated by the primary engine.

In this letter we will first discuss the properties of the bursts. We subsequently compute the neutrino flux inevitably produced when the protons interact with fireball photons. Their observation would provide incontrovertible evidence for the pionic origin of the additional high energy component in the burst and support the speculation that GRB are the sources of the highest energy cosmic rays.

Is a kilometer-scale neutrino telescope such as IceCube sensitive enough to shed light on these questions? High energy neutrinos are produced in the fireball when protons produce pions in interactions with the low-energy photon field:
\begin{equation}
p\,\gamma\rightarrow \Delta^{+}\rightarrow
\left\{
\begin{array}
{lll}n\,\pi^{+}&&\mbox{1/3 of the cases}\\
p\,\pi^{0}&&\mbox{2/3 of the cases}
\end{array}
\right.\,
\end{equation}
The neutral pions decay as $\pi^{0}\rightarrow 2\gamma$, and the charged pions decay as $\pi^{+}\rightarrow\mu^{+}\,\nu_{\mu} \rightarrow e^{+}\,\nu_{e}\,{\nu}_{\mu}\,\nu_{\mu}.$
Here, a single neutrino carries approximately $1/4$th of the $\pi^{+}$ energy and a photon carries $1/2$ of the $\pi^{0}$ energy. The calculation of the neutrino flux \citep{WB} has been performed in detail for the EGRET bursts~\citep{guetta, becker} with the following results: whereas an average burst produces only $\sim 10^{-2}$ neutrinos, bursts that are unusually energetic or nearby produce an observable flux in a kilometer-scale neutrino telescope of order 10 events per year. We suggest that the power-law high-energy spectral feature identifies such bursts.

We will compute the neutrino fluxes expected in IceCube using two methods: the standard fireball model, and the bolometric method which relates the energy in neutrinos from the decay of charged pions to the observed photon energy assuming that it is
of pionic origin. We will conclude that the neutrino rates from all 3 bursts are predicted to be larger than the average, but that a burst like GRB941017 extending to GeV energy will be observed by IceCube. We remind the reader that IceCube observes cosmic neutrinos in a background of neutrinos produced in the atmosphere. Given that neutrinos of GRB origin are relatively energetic and that the direction and time of the events can be correlated to satellite alerts, the atmospheric background is suppressed and a single neutrino may represent a conclusive observation.

\section{Detection of GRBs at high energies\label{extra_component}}
In fireball phenomenology the main contribution to the opacity comes from the annihilation of pairs of photons into $e^{\pm}$ pairs. The observation of a non-thermal spectrum up to $E_{\mbox{\tiny{MAX}}}$ can be used to constrain the minimum bulk Lorentz factor $\Gamma_{\mbox{\tiny{MIN}}}$ required to make the source optically thin. From the condition $\tau_{\gamma\gamma}(E_{\mbox{\tiny{MAX}}})<1$ a lower limit for $\Gamma$ can be derived that requires a minimum value of the jet boost factor proportional to the highest energy $E_{\mbox{\tiny{MAX}}}$ observed:
\begin{equation}
\label{eq:gammaemax}
\Gamma_{\mbox{\tiny{MIN}}}=A(\mbox{z},\delta t,\beta) \cdot \bigg(\frac{E_{\mbox{\tiny{MAX}}}}{m_e c^2}\bigg)^{\Omega(\beta).}
\end{equation}
Here $\beta$ is the spectral index of the high energy part of the Band function and $\delta t$ the variability time of the GRB. Here we use the expressions for $A(\mbox{z},\delta t,\beta)$ and $\Omega(\beta)$ from \citet{Gmin}.

\begin{figure}[h!]
\plotone{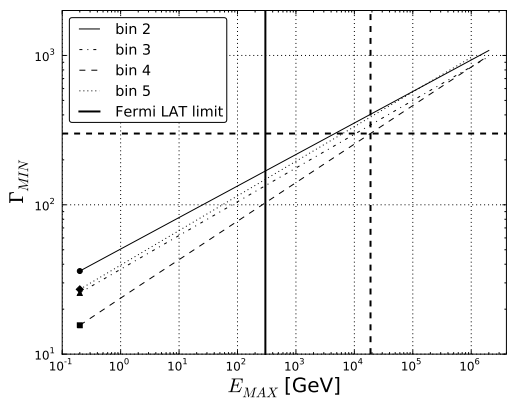}
\caption{$E_{\mbox{\tiny{MAX}}}$ Vs. $\Gamma_{\mbox{\tiny{MIN}}}$ for time bins 2--5 of GRB941017. $E_{\mbox{\tiny{MAX}}}=18.9$\,TeV corresponds to a boost factor of $\sim$\,300, as indicated by the dashed lines. Such an energetic photon would exceed the Fermi LAT detection capability ($\sim$\,300\,GeV).}
\label{fig:gVsEmax941017}
\end{figure}

The first clear detection of a GRB with a power-law component in addition to the standard Band spectral form was made by the EGRET satellite with the observation of GRB941017 in photons with energies up to 200 MeV. The observation being statistics limited, it is of particular importance that EGRET did not detect a cutoff of the power-law component and the flux therefore potentially extended to higher energy\,\citep{hh_04}. We plot $\Gamma_{\mbox{\tiny{MIN}}}$ versus  $E_{\mbox{\tiny{MAX}}}$ for the measured time intervals numbered 2, 3, 4 (Figure~\ref{fig:gVsEmax941017}), described by the parameters of Eq.~\ref{eq:gammaemax}. Here $A=10.2166$, $\Omega=0.211$ for bin 2; $A=6.6721$, $\Omega=0.226$ for bin 3; $A=3.3587$, $\Omega=0.258$ for bin 4, and $A=6.7996$, $\Omega=0.231$ for bin 5. A typical boost factor of $\Gamma\sim$\,300 is compatible with an extension of the high-energy power law to a maximum of 18.9 TeV.

\begin{figure}[h!]
\plotone{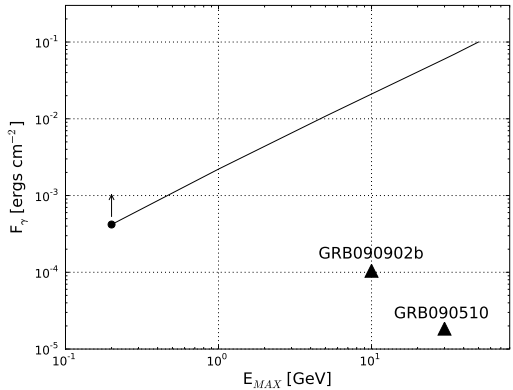}
\caption{Fluence of the high energy component above $30$\,KeV as a function of the maximum energy $E_{\mbox{\tiny{MAX}}}$ for GRB941017 and observed fluences for GRB090510 and GRB090902b.}
\label{fig:fluences}
\end{figure}

The identification of GRB941017 as a potentially powerful neutrino emitter\,\citep{hh_04} suggests a comparison with the similar bursts observed by Fermi. Time integrated spectra of the recently observed GRB090510 and GRB090902b show statistically significant deviations from the Band function fit that extend to $\sim$30\,GeV. In Figure~\ref{fig:fluences}, a comparison between the fluences of the Fermi bursts and the fluence of GRB941017 is presented revealing that the Fermi bursts are significantly less bright, especially if GRB941017 reached higher values of $E_{\mbox{\tiny{MAX}}}$.

Deviations from the Band-only fit in the spectra are particularly interesting in the context of hadronic acceleration within the fireball and relate closely to predictions of neutrino fluxes detectable on Earth with km$^3$-scale telescopes. If the extra high-energy components originate from pionic photons they provide an optimal benchmark for testing models of hadronic acceleration in GRB engines.

\section{Neutrinos\label{fireNus}}

In the hadronic fireball a burst of high-energy neutrinos is expected to accompany the prompt flux of gamma rays. Assuming that electrons and protons are shock-accelerated in the same region, the neutrino spectrum can be calculated from the observed photon Band spectrum using conventional fireball phenomenology. The calculation is routine and is described in detail in~\citet{guetta} and~\citet{ kappes}. Using the observed spectral parameters listed in~\citet{egret94},~\citet{090510} and~\citet{090902b}  the fireball model (Table~\ref{tab:nuParms}) predicts $9.7\times10^{-2}$ neutrinos from GRB941017,  $3.5\times10^{-4}$ neutrinos from GRB090510, and $1.3\times10^{-2}$ neutrinos from GRB090902b in IceCube, using the IC80 neutrino effective area presented in~\citet{effarea}. 

\begin{table}[h!b!p!]
{
\caption{Parameters for fireball neutrino fluxes. The quantity $f_{e}$ in~\citet{kappes} is defined as $\epsilon_{p}/\epsilon_{e} = 1/f_{e}$.}
\label{tab:nuParms}
}
{
\begin{center}
\begin{tabular*}{0.9\columnwidth}{c|c|c}
  & \small{GRB090510} & \small{GRB090902b} \\[3pt]
\hline \hline
\small{$\epsilon_{\gamma}$}   & \small{$2.771$ MeV} & \small{$0.726$ MeV} \\
\small{$\alpha_{\gamma}$}    & \small{0.58} & \small{0.61}\\
\small{$\beta_{\gamma}$}  & \small{2.83} & \small{3.8}\\
\small{$\langle x_{p\rightarrow\pi}\rangle$}  & \small{0.2} & \small{0.2}  \\
\small{$\epsilon_{p}/\epsilon_{e}$} & \small{15.6}  & \small{45.6} \\
\small{$\Gamma$}  & \small{1260} & \small{1000} \\
\small{$\alpha_{\nu}$}   & \small{0.17} & \small{-0.8}\\
\small{$\beta_{\nu}$}    & \small{2.42} & \small{2.39}\\
\small{$\gamma_{\nu}$}    & \small{4.42} & \small{4.39}\\
\small{$\epsilon_{\nu\,1}$}  & \small{$1.18\times10^6$ GeV} & \small{$1.30\times10^6$ GeV} \\
\small{$\epsilon_{\nu\,2}$}   & \small{$3.59\times10^8$ GeV} & \small{$5.29\times10^8$ GeV} \\
\small{z}  & \small{0.903} & \small{1.822} \\
\small{$t_{\mbox{\tiny{var}}}$}  & \small{0.01 s} & \small{0.053 s} \\
\small{$T_{90}$}  & \small{2.1 s} & \small{21.9 s} \\
\end{tabular*}
\end{center}
}
\end{table}

\begin{figure}[h!]
\plotone{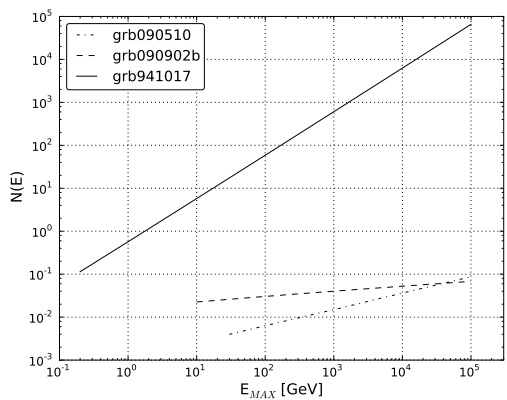}
\caption{Neutrino events associated with the power-law spectral component from each burst as a function of $E_{\mbox{\tiny{MAX}}}$.}
\label{fig:bolo_events}
\end{figure}

We also expect that if the observed power-law components of the photon flux result from the decay of neutral pions photoproduced by protons on fireball photons, then there should also be a related flux of neutrinos due to the production and decay of charged pions. However, the observed power-law spectral components are quite flat ($E^{-1}-E^{-1.6}$) and do not follow the approximate $E^{-2}$ spectral dependence expected for shock acceleration. This spectral behavior is compatible with synchrotron radiation from highly relativistic electrons and positrons, which are produced when high-energy photons scatter in the fireball photon field. While the spectral behavior is modified in this cascading
process, the total energy contained in the observed photons corresponds to the total energy originally going into high-energy photon production. Hence, we can assume that the fluence of gamma rays $F_{\gamma} \left(\equiv  \int E\, \phi(E)\, dE \,\right)$ remains constant. Moreover, since both neutrinos and gamma rays originate in pions, we know that their energy fluences must be proportional $F_{\nu} \propto F_{\gamma}$. If we assume that each gamma ray takes one half of the pion energy and each neutrino takes one quarter, and that since there are two neutral pions per charged pion there are four gamma rays per neutrino (after oscillations), we find the proportionality constant to be approximately 1/8~\citep{bolo}. Therefore, if we recalculate the gamma-ray fluences from the power-law spectral components for different values of $E_{\mbox{\tiny{MAX}}}$, we can equate those fluences to the neutrino fluence and hence, given the spectrum of neutrinos from pion decay, calculate the normalization of the neutrino flux. Using the IceCube effective area, we find the total number of neutrinos detected in IC80 from each burst as a function of $E_{\mbox{\tiny{MAX}}}$ (Figure~\ref{fig:bolo_events}). We call attention to the fact that a burst with the same parameters as GRB941017, extending to  $E_{\mbox{\tiny{MAX}}}>\!2$ GeV, will produce $>\!1$ neutrino event in IceCube. Given the absence of background events over such a short time interval, a single event would be statistically significant and would constitute discovery of proton acceleration in GRBs. 

\section{Summary and Conclusions}
Three gamma-ray bursts with statistically significant high-energy power-law spectral components have been detected thus far. The observed characteristics of the bursts indicate (Eq.~\ref{eq:epse}) that it is likely that a large fraction of their total energy is hadronic rather than electromagnetic. As a result, we assume that these bursts are accelerating protons and calculate the associated fluxes of neutrinos. While the event rates of neutrinos derived from standard fireball phenomenology are small, we take the existence of the high-energy power-law components as indicative of the decay of $\pi^{0}$-mesons. This allows us to calculate the magnitude of the neutrino flux from the related charged pions, and we find that a burst like GRB941017 will be observable in IceCube if its power-law component extends to energies in excess of $\sim1$ GeV.  

\acknowledgments
J.K.B. and M.O. acknowledge support from the Research Department of Plasmas with Complex Interactions (Bochum). F.H. and A.\'O.M were supported in part by the National Science Foundation under Grant No.~OPP-0236449, in part by the
U.S.~Department of Energy under Grant No.~DE-FG02-95ER40896, and in part by the University of Wisconsin Research Committee with funds granted by the Wisconsin Alumni Research Foundation.



\begin{thebibliography}{}

\bibitem[Abbasi(2009) ]{kappes}
Abbasi, R. \textit{et al.}, \apj \, \textbf{701}, 1721 (2009)

\bibitem[Abdo(2009) ]{090510}
Abdo, A.A. \textit{et al.}, arXiv.0908.1832 (2009)

\bibitem[Ahlers(2005) ]{ahlers}
{Ahlers}, M. \textit{et al.}, \prd\, \textbf{72}, 023001(2005)

\bibitem[Alvarez-Mu\~{n}iz(2002) ]{bolo}
Alvarez-Mu\~{n}iz, J. \textit{et al.}, \apj \, \textbf{576}, L33 (2002)

\bibitem[Alvarez-Mu\~{n}iz(2004) ]{hh_04}
Alvarez-Mu\~{n}iz, J. \textit{et al.}, \apjl \, \textbf{604}, L85 (2004)

\bibitem[Becker(2006) ]{becker}
Becker, J.K. \textit{et al.}, Astroparticle Physics \, \textbf{25}, 118 (2006)

\bibitem[Bissaldi(2009) ]{090902b}
Bissaldi, E. \textit{et al.}, arXiv:0909.2470 (2009)

\bibitem[Gonz{\'a}lez(2003) ]{egret94}
{Gonz{\'a}lez}, M. M. \textit{et al.}, \nat\, \textbf{424}, 7491 (2003)

\bibitem[Gonzalez-Garcia(2009) ]{effarea}
Gonzalez-Garcia, M.C. \textit{et al.}, Astroparticle Physics \, \textbf{31}, 437 (2009)

\bibitem[Guetta(2004) ]{guetta}
Guetta, D. {\it et al.}, Astroparticle Physics \textbf{20}, 429 (2004)

\bibitem[Lithwick(2001) ]{Gmin}
Lithwick, Y. and Sari, R., \apj\, \textbf{555}, 540 (2001)

\bibitem[Vietri(1995) ]{vietri_1995}
Vietri, M., \apj\, \textbf{453}, 883 (2005)

\bibitem[Waxman(1995) ]{waxman_1995}
Waxman, E., \prl\, \textbf{75}, 386 (1995)

\bibitem[Waxman(1998) ]{WB}
Waxman, E. and Bahcall, J., \prd\, \textbf{59}, 23002 (1998)

\end{thebibliography}
\end{document}